\documentclass[12pt]{aastex}

\newcommand{\ms}{\mbox{m s$^{-1}~$}}

\newcommand{\mjupe}{M$_{\rm JUP}$}

\newcommand{\msini}{$M \sin i~$}

\newcommand{\etal}{{\it et al.}}
\newcommand{\eg}{{\it e.g.}}
\newcommand{\ie}{{\it i.e.}}

\newcommand{\au}{{\sc au}}

\received{}
\accepted{}

\slugcomment{Submitted to ApJ: 12, Nov. 2001}

\shortauthors{Laughlin, Chambers \& Fischer}
\shorttitle{Dynamics of the 47 UMa System}
\begin{document}

\title{A Dynamical Analysis of the 47 UMa Planetary System$^{1}$}
\author{Gregory Laughlin \altaffilmark{1,2}, 
John Chambers \altaffilmark{2},
Debra Fischer \altaffilmark{3}} 

\email{laughlin@ucolick.org}

\altaffiltext{1}{Department of Astronomy, UCO/Lick Observatory,
UCSC, Santa Cruz, CA USA  95064}

\altaffiltext{2}{NASA Ames Research Center
MS 245-3, Moffett Field, CA 94035}

\altaffiltext{3}{Department of Astronomy, University of California,
Berkeley, CA USA  94720}

\begin{abstract}

Thirteen years of Doppler velocity measurements have revealed the
presence of two planets orbiting the star 47 Ursa Majoris on low
eccentricity orbits.  A 2-Keplerian fit to the radial velocity data
suggests that the inner planet has a period $P_b = 1089.0 \pm 2.9$ d,
and a nominal ($\sin(i)=1$) mass \msini = 2.54 \mjupe, while the outer planet has a
period $P_c = 2594 \pm 90$ d, and a mass \msini = 0.76 \mjupe. These
mass and period ratios suggest a possible kinship to the
Jupiter-Saturn pair in our own solar system.  We explore the current
dynamical state of this system with numerical integrations, and
compare the results with analytic secular theory. We find that the
planets in the system are likely participating in a secular resonance
in which the difference in the longitudes of pericenter librates
around zero. Alternately, it is possible that the system is participating
in the 7:3 mean motion resonance (in which case apsidal alignment does
not occur). Using a
self-consistent fitting procedure in conjunction with numerical
integrations, we show that stability considerations restrict the
mutual inclination between the two planets to $\sim$ 40 degrees or less, and
that this result is relatively insensitive to the total mass of the
two planets. We present hydrodynamical simulations which measure the
torques exerted on the planets by a hypothesized external
protoplanetary disk. We show that planetary migration in response to
torques from the disk may have led to capture of the system into a 7:3
mean-motion resonance, although it is unclear how the eccentricities
of the planets would have been damped after capture occured.  We show
that Earth-mass planets can survive for long periods in some regions
of the habitable zone of the nominal co-planar system.  A set of
planetary accretion calculations, however, shows that it is unlikely
that large terrestrial planets can form in the 47 UMa habitable zone.
\end{abstract}


\section{Introduction}
\label{intro}

Planetary systems with multiple gas giant planets in long-period
circular orbits would mark the long-sought true analogues of our Solar
System. The first extrasolar multiple planet system was discovered
around Upsilon Andromedae (Butler et al. 1999) and since then, pairs
of planets have been detected around HD83443 (Mayor et al. 2000),
HD168443 (Marcy et al. 2001a, Udry et al. 2000), GJ876 (Marcy et
al. 2001b), and most recently, 47 UMa (Fischer et al 2002). Radial
velocity trends among the known planet-bearing stars suggest that
additional planetary companions may ultimately be found in more than
half of the stars with one detected planet (Fischer et al. 2001).

When first announced in 1996, thirty radial velocity measurements
suggested a single 2.5 \mjupe\ planet in a circular 2.2 \au\  orbit around
47 UMa (Butler \& Marcy 1996).
As described in Fischer et al 2002, 61 subsequent radial
velocity measurements (for a total of 91 velocities) suggest
that a second, less massive planet also exists. The nominal mass
of this second planet is 0.76 \mjupe, its eccentricity is small, and
its semi-major axis is roughly 3.7 \au. The current best-fit properties
of the system (as reported in Fischer et al 2002) are shown in Table 1.

Of all the multiple systems found to date, 47 UMa is by far the most
reminiscent of our own Solar System. The star 47 UMa is similar
in mass, age and rotational velocity to the Sun, with  $M=1.03 M_{\odot}$, 
$\tau = 7 {\rm Gyr}$, and $P_{rot}=24 {\rm d}$ (Noyes et al 1984,
Baliunas et al. 1995). The accumulated radial velocity measurements
suggest that 47 UMa bears a startling resemblance
to the Jupiter-Saturn pair in our own solar system. The orbital period
ratios for both systems are quite similar
(2.38:1 for 47 UMa, 2.49:1 for Jupiter-Saturn), and to within the
radial velocity measurement errors, the mass ratios are identical (3.34:1 for
co-planar configurations of 47 UMa versus 3.34:1 for Jupiter-Saturn).  The
major difference between the two systems is one of overall scale. The
orbital period of 47 UMa b is less than one-fourth that of Jupiter,
and the nominal mass is $\sim 2.5$ times greater. We thus expect that
the planets in the 47 UMa system will experience more significant
mutual perturbations than do Jupiter and Saturn, and on a shorter
timescale. A dynamical analysis of the properties of the 47 UMa system
constitutes the main goal of this study.  In particular, we would like to
know how dynamical arguments can constrain the parameter space allowed
to the system by the radial velocity observations, and
we would like
to gain a better idea of whether the apparent kinship between the
solar system and 47 UMa goes beyond a superficial resemblance.

\section{Numerical Integrations of the 47 UMa System}

Because the outer planet in the 47 UMa system has made only two full
revolutions since the beginning of the radial velocity observations,
the eccentricity, $e_{c}$ of the outer planet is not well determined.
In fact, $e_{c}=0.3$ provides almost as good a fit to the radial
velocity data as does $e_{c}=0.005$.
Furthermore, the inclinations of
the orbits to the plane of the sky are unknown, and there are therefore
many different configurations of the system which are consistent with
the current radial velocity dataset.

One can get an idea of which configurations are allowed for the 47 UMa
planets by performing numerical integrations of the star -- 2 planet
system, and in particular, we can determine which ranges of mass
factors, mutual inclinations and outer planet eccentricities are both
consistent with the radial velocity dataset and are also dynamically
stable. This approach has met with considerable success in 
constraining the properties of the multiple planet system orbiting
Upsilon Andromedae (see e.g. Laughlin \& Adams 1999, Rivera \& Lissauer 2000,
Lissauer \& Rivera 2001,
or Barnes \& Quinn 2001, or Gozdziewski et al 2001).

In a preliminary dynamical analysis performed by Fischer et al 2002, it
was found that during the course of an individual trial integration, the 
periodic maxima in the 
osculating eccentricity of the outer planet provide a good running
diagnostic of the distance of the system from instability. Once the
eccentricity of the outer planet ventures above $e_{c}=0.2$, the
destruction of the system (via the near-parabolic ejection of the outer 
planet) is gererally imminent. \footnote{The starting conditions for the
simulations reported by Fischer et al 2002 employed a preliminary version
of the 2-planet fit to the 47 UMa radial velocity data set. This fit
differed slightly from the osculating elements of Table I (which is the
same as the published elements in Fischer et al 2002). The exact osculating
elements used in Fischer et al (2002) were: $P_{b,c} (\rm d)=1090.5,
2560.7$, ${T_{peri}}_{b,c} (\rm d)=2452523.9, 2451293.7$, $e_{b}=0.06$,
$\varpi_{b,c} ({\rm deg})=167.5, 119.0$, $K_{b,c} ({\rm m/s})=49.4, 11.0$,
${\rm epoch}={\rm JD 2451293.7}$.}
Table 2 shows the outcome of a grid of
calculations done for various co-planar configurations of the
fit to the system listed in Table 1
(with the co-planar ``mass factor'' given by ${\sin (i)}^{-1}$, where
$i$ is the inclination of the system with respect to the plane of the sky), and
with various values for the $e_c$. All simulations were started at epoch
JD 2451363.5, the moment of periastron passage for planet c.
In each case, the simulation was
run for either 100 million years, or until instability occured (listed
as ``unst''). For stable configurations, we list the largest
eccentricity obtained by the outer planet during the course of the
simulation. It is clear that for stability, $e_{c}<0.2$, but that for
small values of $e_c$, the system planets can be up to six times
heavier than their nominal masses, and yet still be dynamically
stable. We note that astrometric limits from the Hipparcos satellite
rule out mass factors greater than 3 (Fischer et al 2002).

We can compare the results of the integrations with an analytic
stability criterion. In the 3-body problem, certain
values of the energy $E$ and angular momentum $L$ make it impossible
for close encounters to occur between the outer two bodies in the
system. Hence in the 47 UMa system, close encounters between the two
planets can only occur for certain ranges of the initial orbital
elements. In particular, collisions between planets are prevented if 
\begin{equation}
-\left(\frac{2m_{\rm tot}}{G^2m_{\rm pair}^3}\right)L^2E>S_{\rm crit}
\label{eq_stable}
\end{equation}
where $m_{\rm tot}=m_\star+m_b+m_c$ is the total mass of the system,
$m_\star$ is the mass of the star, $m_{\rm
pair}=m_\star m_b+m_\star m_c+m_bm_c$, and
\begin{equation}
S_{\rm crit}=1+\frac{3^{4/3}m_bm_c}{m_\star^{2/3}(m_b+m_c)^{4/3}}
-\frac{m_bm_c(11m_b+7m_c)}{3m_\star(m_b+m_c)^2}+\cdots
\end{equation}
(Marchal and Bozis 1982, Gladman 1993).

When the planets are not in a close encounter, we can express $E$
and $L$ approximately in terms of the osculating orbital elements of
their Keplerian orbits about 47 UMa. Equation 1 now becomes
\begin{equation}
\frac{(1+\alpha\mu)}{(1+\mu)^{3}}\left[\left(1-\frac{e_b^2}{2}\right)
+\mu\alpha^{-1/2}\left(1-\frac{e_c^2}{2}\right)\right]^2
>S_{\rm crit}
\label{eq_stable2}
\end{equation}
where $\mu=m_c/m_b$ and $\alpha=a_b/a_c$. Equation 3 assumes small
eccentricities for both planets, and $m_b, m_c \ll 1$.

For the case in which the planets have coplanar orbits inclined at
an angle $i$ to the sky, $S_{\rm crit}$ is given by
\begin{equation}
S_{\rm crit}=1+0.01616(\sin i)^{-2/3}+0.00182(\sin i)^{-1}+\cdots
\end{equation}
for the values of $K$ appropriate to the 47 UMa system given in Table~1.
Note that the lefthand side of Eqn 3
is independent of $i$ at this level of approximation.

We have applied the analytic stability criterion to the numerical
integrations shown in Table~2. The simulations above the ragged
line are prevented from undergoing planetary close encounters
(that is, they are ``Hill stable'') by conservation of energy and
angular momentum. The simulations below the line are not prevented
from having close encounters according to the analytic criterion,
although other mechanisms may maintain stability in practice.

The results of the integrations show some interesting disagreements
with the analytic
criterion. The threshold for instability tends to occur at lower
values of $e_c$ for increasing masses of the planets (decreasing
$\sin i$). For low values of $e_c$, some of the high-mass systems
are stable despite the fact that they are not constrained to be
so by the analytic criterion. It is possible that other protection
mechanisms are occurring in these cases,
but it is also possible that these systems are
unstable on very long timescales. As a point of curiosity, we
note that the system represented by the table entry having
$\sin(i)^{-1}$=1.75, $e_{c}=0.1$ develops instability within several 
million years. This point in the phase space of initial conditions is
part of a narrow gulley of instability that eats into the table
when $\sin(i)^{-1} \approx 1.75$.

The systems with high $e_c$ are also intriguing. Some of these
systems are unstable despite the fact that the stability criterion
predicts the planets should not experience close encounters. A closer
examination of these integrations reveals that this prediction
is correct. For example, in an integration with $e_c=0.25$ and $\sin i=1.0$, 
the planets never passed within 0.5 \au\ of each other (equal to about
2.5 Hill radii of the inner planet), although the outer planet was
ejected from the system after 40,000 years. In this case, distant
encounters and indirect perturbations (the effect of the inner planet
on the motion of the star) are responsible for the ejection of
the outer planet. 
This system and the other unstable cases above the line in Table~2
are examples of ``Lagrange instability'' in which close encounters
cannot occur, but escape is still possible. We note that analytic
stability criteria similar to Eqn~1 are widely used in the study
of hierarchical 3-body systems, but in general these criteria cannot
establish whether a system is Lagrange stable. As our simulations show,
planetary systems that are close to the Hill stability limit can 
be Lagrange unstable on short timescales, so one must be careful
when applying the Hill stability criterion.

It is also worthwhile to point out that the integrations in Table
2 do not represent the last word on the stability of the system. 
The orbital elements in Table 1 all have associated uncertainties,
and if alternate initial conditions are drawn from the 1-$\sigma$
distributions associated with the parameters, then the stability
boundary is subject to change, sometimes with extraordinary sensitivity.
Furthermore, the development of
instability is a statistical rather than a deterministic process.
If stability is observed over a particular interval for a particular
simulation, it does not prove that the configuration is unconditionally
stable. Numerical experiments of the sort shown in Table 2 must
therefore be viewed from a qualitative rather than quantitative 
standpoint.

Figure 1 shows three sets of numerical integrations. Two are of
nominal co-planar configurations of the 47 UMa system with $e_c=0.005$
and $e_c=0.15$ respectively. The third integration is of the
Jupiter-Saturn pair in our own solar system. The time evolution of the
orbital eccentricities and relative longitude of periastron $\Delta
\varpi$ are indicated by the red curves in the figure. In each system,
the eccentricities of the planets oscillate in antiphase, as they
exchange orbital angular momentum. For 47 UMa in the low $e_c$ case,
the planets lie in a secular resonance in which
$\Delta\varpi=\varpi_2-\varpi_1$ librates about zero.\footnote{
Throughout this paper we use the term secular resonance to
mean a resonance in which the secular argument $\varpi_{2} - \varpi_{1}$
is librating. This should not be confused with the secular
resonance in which two of the eigenfrequencies of the Laplace-Lagrange
secular Hamiltonian coincide (Kinoshita and Nakai 2000).}
This is a stable
configuration in which the major axes of each orbit remain roughly
aligned, preventing close approaches between the planets. For Jupiter
and Saturn, and for 47 UMa in the high $e_c$ case, $\Delta\varpi$
circulates through 360 degrees. This periodically brings the major
axes of the orbits into antialignment, at which point the
perturbations between the planets are strongest. In these cases shown
in Figure 1, the system remains stable, but larger initial values of
$e$ lead to instability (see Table 2).

The black curves in Figure 1, show the results of applying classical
Laplace-Lagrange secular theory to the same systems. In this theory,
the mutual planetary perturbations are expanded as a Fourier series
in $e$. Short-period terms, mean-motion resonances, and terms
$O(e^4)$ and higher are all neglected, leaving a set of coupled linear
differential equations. Following Murray and Dermott (1999), the
solution to these equations can be expressed as
\begin{eqnarray}
e_1\sin\varpi_1&=&e_{11}\sin\psi_1 + e_{12}\sin\psi_2 \nonumber \\
e_1\cos\varpi_1&=&e_{11}\cos\psi_1 + e_{12}\cos\psi_2 \nonumber \\
e_2\sin\varpi_2&=&e_{21}\sin\psi_1 + e_{22}\sin\psi_2 \nonumber \\
e_2\cos\varpi_2&=&e_{21}\cos\psi_1 + e_{22}\cos\psi_2
\label{seceq1}
\end{eqnarray}
with $\psi_i=g_it+\beta_i$, where $t$ is the time and $e_{ji}$, $g_i$
and $\beta_i$ are constants determined by the masses and initial
orbital elements of the planets. The subscripts 1 and 2 refer to the
inner and outer planets, respectively.

The linear secular theory qualitatively reproduces the form of the
evolution found in the numerical integrations. It also predicts the
amplitude of the oscillations in $e$, and correctly determines whether
or not the planets are librating about the secular resonance.
However, the secular periods predicted by the linear theory are
wrong. In the case of Jupiter and Saturn, this difference is largely
due to the 5:2 near mean-motion resonance between these planets, which
is neglected in the secular theory (Murray and Dermott 1999).  
The discovery of this near-resonance allowed Laplace (1785; see also 
Laskar 1996)
to identify an additional 900 year periodicity in the orbital elements
of Jupiter and Saturn, which is visible in Figure 1 as a high-frequency
modulation in Saturn's eccentricity.
Conversely, for the specific set of osculating orbital elements given
in Table 1, the planets in
the 47 UMa system are not close to any strong resonances (the nearest
is the 7:3), and the shortcomings of the secular theory are probably
caused by neglect of higher order terms in the masses and
eccentricities.

Despite its limitations, we can use the Laplace-Lagrange theory to
understand why the 47 UMa planets are in a secular resonance for small
$e_c$ but not at larger values, and why Jupiter and Saturn do not lie
in the same resonance. Rearranging Eqns. 5, we get
\begin{equation}
e_1e_2\cos\Delta\varpi=(e_{11}e_{21}+e_{12}e_{22})
+(e_{11}e_{22}+e_{12}e_{21})\cos(\psi_1-\psi_2)
\end{equation}

Since $\psi_1-\psi_2$ increases linearly with time, the behavior
of $\Delta\varpi$ will depend on a quantity $S$ given by
\begin{equation}
S=\left|\frac{e_{11}e_{22}+e_{12}e_{21}}{e_{11}e_{21}+e_{12}e_{22}}\right|
\end{equation}

When $S>1$, $\Delta\varpi$ can take any value and the planets lie
outside the secular resonance with $\Delta\varpi$ circulating.  When
$S<1$, $\Delta\varpi$ is restricted to a particular range of values,
and the planets lie in the resonance. In the latter case, librations
can take place about 0 (major axes aligned) or $180^\circ$ (major axes
antialigned).

Figure 2, shows the extent of the libration
regions for the planets in 47 UMa, and for Jupiter and Saturn, as
a function of the initial values of $\Delta\varpi$ and $e$ of the
outer planet. In the case of 47 UMa with low $e_c$, the planets lie
deep inside the secular resonance, with $\Delta\varpi$ librating about
0. However, when $e_c$ is larger (for the same initial value of
$\Delta\varpi$), the system lies outside the resonance.
Jupiter and Saturn also lie outside the resonance, but relatively
small changes in Saturn's eccentricity or $\Delta\varpi$ would bring the
planets into resonance. (The proximity to the secular resonance is
apparent from the form of the $\Delta\varpi$ curve for Jupiter and
Saturn in Figure 1.)

Of the multiple planet systems known to date, two others (GJ 876 and
$\upsilon$ Andromedae) are also likely to be involved in the secular 
resonance. In both of these systems, the libration amplitudes of the
outer planets appear to be small (Laughlin \& Chambers 2001,
Chiang, Trebachnik, \& Tremaine 2001), hinting at some sort of
dissipative evolution following resonance capture. 
By contrast, for the particular parameter values listed in Table 1,
47 UMa is undergoing very large librations around the resonance, and
Jupiter and Saturn are out of the resonance. 
Given the current orbital uncertainties for 47 UMa, it is not clear
whether this difference is significant, or whether the existence of the
possible secular resonance in 47 UMa, and the circulation in the Jupiter-Saturn
system is simply a matter of chance.

\section{Mutually Inclined Configurations}

The Doppler radial velocity technique measures only the line-of-sight
velocity of the star, and is thus unable to determine the inclinations
and nodes of the planetary orbits if the orbits are assumed to be Keplerian.
In systems such as GJ876 (Marcy et al 2001b), in which the planets are 
strongly interacting, and where the orbital periods are short enough so
that the system can be observed over many cycles, fits to the full N-body
motion can potentially determine all of the orbital parameters of the
system (Laughlin \& Chambers 2001, Rivera \& Lissauer 2001). In the
47 UMa system, however, the orbit of the outer planet has been followed
with 3--5 m/s precision for only a single orbital period. It is presently
impossible to obtain the inclinations and nodes from the radial velocity
curve itself. Numerical integrations of allowed system configurations
can, however, place limits on the system.

At a given epoch, the orbit of a planet in a multi-planet system is
parameterized by six orbital elements ($a$, $e$, $i$, $\omega$,
$\Omega$, and $l$), along with a mass, m. Here, the inclination angle
$i$ is understood to be with respect to the plane of the sky. For a
system model containing two planets, the overall invariance of the
radial velocity curve with respect to system rotations around the line
of sight can be accounted for by specifying a parameter $\Omega_{12}$
corresponding to the relative difference between the lines of nodes
for the two planets, yielding, for two planet systems, a total of 13
system parameters. When fitting to a radial velocity curve, a final
parameter corresponding to the velocity zero point is also required.

We generate mutually inclined fits to the 47 UMa radial velocity data
by randomly selecting fixed values for the inclinations, $i_{1}$ and
$i_{2}$, and the angular separation of the lines of nodes,
$\Omega_{21}$. The remaining eleven parameters are then obtained by
using a Levenberg-Marquardt minimization scheme driving a three-body
integrator to obtain a low $\chi^{2}$ fit to the radial velocities (as
described in Laughlin \& Chambers, 2001). This fitting procedure
automatically accounts for the dependence of $\omega_{i}$ on viewing
geometry (see Chiang, Tabachnik, \& Tremaine, 2001). Each allowed
configuration is then integrated for $10^{6}$ years to check
stability. In Fischer et al (2002), 
and in section 2 above, we found that configurations of
the 47 UMa system in which the osculating eccentricity of the outer
planet exceeds $e_{2}=0.2$ are generally dynamically unstable on 100
million year timescales, whereas systems in which $e_{2}<0.2$ are
generally stable.

In Figure 3, we plot the results of the orbital integrations of 150 
mutually inclined fits to the radial velocity data. The three-dimensional
parameter space spanned by $i_1$, $i_2$, and $\Omega_{21}$ is projected onto a 
two-dimensional space spanned by the mutual inclination, 
$\cos i_{12}=\cos i_{1} \cos i_{2} + \sin i_{1} \sin i_{2} \cos \Omega_{21}$,
and the total planetary mass, $m_{1}+m_{2}$. 
This telescopes the three parameters down to two for
easier interpretation. If a system reaches ${e_2}_{max}>0.4$, it is plotted 
as a filled circle. Systems that reach $0.2< {e_2}_{max} <0.4$ are plotted
as stars, and stable systems are plotted as open circles.
The figure shows an interesting result that mutual inclination must be
less than 45 degrees for the system to be stable, and this result is
surprisingly independant of the total mass in the system. 

\section{Hydrodynamical Simulations}

The mass and period ratios of the 47 UMa planets are reminiscent of
the Jupiter-Saturn pair in our own solar system. Nevertheless, the 47
UMa system packs at least 3.5 Jupiter masses into a region which was
within the so-called ``snow-line'' of 47 UMa's
protoplanetary nebula, where temperatures were below 150K, and where
water ice grains would have survived evaporation at the time when 
planetesimals were forming. It thus seems
reasonable to imagine that the 47 UMa planets formed in a colder
environment, where icy materials were available to build
protoplanetary cores, and then migrated inwards. 

The phenomenon of planet-disk interaction and migration is now becoming
better understood. In early phases, when the mass of a planet is small,
an orbiting planet induces a linear spiral response in its parent disk. 
The spiral pattern in the disk material lying inside the planet's orbit exerts a
positive torque on the planet, which causes outward migration, whereas the
spiral pattern in the outer disk exerts a negative torque on the planet
and causes inward migration.

If the mass of the planet, $M_{pl}$, becomes large enough 
so that the Roche radius,
$$R_{R}=({M_{pl}\over{3 M_{\star}}})^{1/3} a_{pl}$$
is larger than the local scale height $h$ of the disk, then the migrating
planet can open up a gap if the disk has a sufficiently low viscosity:
$$\alpha_{disk} \le \alpha_{max}={1\over{40}}{M_{pl}\over{M_{\star}}}({a_{pl}\over{h}})^{2}.$$
In a system where two planets have a semi-major-axis ratio of
$a_{2}/a_{1} \sim 2$, work by Bryden et al (2000), and Kley (2000)
shows that the ring
of gas between the two planets is removed within several hundred orbital
periods. Initially, disk material would be present interior
to the orbit of the inner planet. The positive torque exerted by this
inner disk is likely to cause outward migration of the inner planet,
while the inner disk itself is gradually depleted by accretion onto the
central star.  This means that the nascent 47 UMa system   
may have experienced a phase in which the two planets were present with
effectively their present masses, and were surrounded by an external
disk with little disk mass interior to the inner planet. 
We can therefore ask: would the planets then experience interactions
which would lead to migration, resonance capture or eccentricity pumping?
Can the present state of the system put interesting limits on the amount 
of migration that might or might not have occured?

We have modeled a hypothetical forerunner to the 47 UMa system 
using a 2-dimensional hydrodynamical 
code described by Laughlin (1994). In the simulation,
two planets, with masses equal to the nominal masses in Table 1,
are placed in circular orbits having an initial semi-major axis ratio,
$a_{2}/a_{1}=1.8$. The planets are allowed to interact gravitationally
with an external circumstellar disk having a total mass of 1 $M_{jup}$.
Most of the mass of the circumstellar disk lies beyond the orbit of the
second planet;
the initial surface density profile of the disk has a split-gaussian
profile with
$$\sigma(r)=\sigma_{0} e^{-(r-R_{0})^{2}/w_{i}}$$ for $r<R_{0}$,
and
$$\sigma(r)=\sigma_{0} e^{-(r-R_{0})^{2}/w_{o}}$$ for $r>R_{0}$,
with $w_{i}=1.0$ \au, $w_{0}=2$ \au, and
$R_{0}=6$ \au.

The disk is modeled with a locally isothermal equation of state, in which the
scale height of the disk, $h/r=0.03$, is assumed constant.
The inner and outer radial 
boundaries of the hydrodynamical simulation are located at $R_{in}=1$ \au\
and at $R_{out}=10$ \au. 
The initial planetary orbital radii
lie at $r=2.0$ \au\, and $r=3.6$ \au. Although both planets are immersed in the
disk, they are not allowed to accrete material which passes within
their Roche lobes; the initial disk profile assumes that the outer planet
has nearly completed the gap-opening process. The planets are given
a gravitational softening length $r_{s}=0.30$ \au\, in order to 
minimize erratic local forces arising from individual grid cells. Because
of this large softening length, and because we are modeling a phase 
following gap clearing, material within the Roche lobes of the planets
is not given special treatment. The planet
and the star are integrated using a 4th order Runge-Kutta method with a
timestep set at 1/20th of the CFL condition which limits
the hydrodynamic timestep. A kinematic viscosity
of order $\nu=8. \times 10^{-5} {{\rm AU}^{2} /{\rm yr}}$ is used. 
Mass which flows across the inner
boundary of the hydrodynamical domain is added to the central star.

The simulation employs 256 logarithmically spaced radial zones and
256 evenly spaced azimuthal zones, and was run for 250 orbits of the inner
planet. Figure 4 shows an image of the system at the end of the simulation.
The system has settled down to a 
quasi-steady state configuration in which the outer planet has set up a
trailing spiral response in the disk. Because the outer Lindblad
resonance of the inner planet lies in a region of low surface density,
there is little coupling between the disk material and the inner planet.
The top panel in Figure 5 shows the azimuthal component of the accelerations 
of the two planets due to the disk material. After an initial
period of readjustment, which is associated with the transient response of 
the equilibrium disk to the introduction of the two planets, the disk is
seen to exert a constant torque on the outer planet, causing it to
experience an azimuthal acceleration of $-1.2 \times 10^{-5} {{\rm AU}^{2} /{\rm yr}}$. 
This causes the outer planet to spiral in on a timescale 
${a \over{\vert {\dot a} \vert}} \propto 10^{5} {\rm yr}$
(as shown in the second panel from the top of Figure 5). 
This timescale is consistent with the analytic result that the migration
timescale is the viscous timescale of the disk, and this agreement
has been previously noted in the simulations of, e.g., Bryden et al (2000),
and Kley (2000).
The net azimuthal torque on the inner planet averages to near zero. The
inner planet thus shows very little change
in semi-major axis (second panel from the bottom of Figure 5). During
this independent phase of migration, the eccentricities of both planets
remain near zero. The kinematic viscosity 
$\nu=8.0 \times 10^{-5} {{\rm AU}^{2}/{\rm yr}}$ used in the simulation
corresponds to an ``alpha'' coefficient (Shakura \& Sunyaev 1973)
$\alpha = {\nu \Omega \over{c_{s}^{2}}} = 0.011$ at the radial
location of the outer planet. For the disk under consideration, $\alpha$ 
must be less than $\alpha_{max}=0.02$ for the outer planet to be able to 
maintain a gap. The surface density distribution of the disk (Figure 4)
shows that the outer planet is indeed maintaining a gap in the disk,
although some material is lingering in the vicinity of the stable
Lagrange points $L_{4}$ and $L_{5}$. The larger mass of the inner planet 
allows it to easily maintain its gap. A low mass ring of material
between the two planets is replenished by gas flowing through the gap
maintained by the outer planet.

The result of the hydrodynamical simulation suggests that migration
of the outer planet occurs at a steady rate, and that the disk torque
does not directly couple to the inner planet. We can therefore mimic
the effects of the full gas-dynamical calculation by performing much
less time-consuming 3-body integrations in which a steady azimuthal
acceleration $a_{\phi}=-1.2 \times 10^{-5} {\rm AU}^{2}/{{\rm yr}}$ of magnitude
suggested by the simulations is applied to the outer planet. This 
approach has been applied by Lee and Peale (2002) to study the evolution
of the GJ 876 system, which is now participating in the 2:1 resonance.
Additional simulations of the GJ 876 configuration by Snellgrove,
Papaloizou \& Nelson (2001) have studied resonant capture into the
2:1 resonance using both torqued three body and hydrodynamical approaches.

Figures 6 and 7 show the result of one such simulation. Figure 6 is
a surface of section in which
the ratio of the planets' semi-major axes, $a_{c}/a_{b}$, 
is plotted against the orbital
phase of the inner planet each time the outer planet reaches aphelion.
Mean-motion resonances are visible in this diagram as discrete islands
of points
In the initial phases of the evolution, the outer planet migrates inward
towards the inner planet. Once the planets are locked in resonance, the
quantity $a_{c}/a_{b}$ stops decreasing.
In the case shown, the two planets skirt
the 5:2 resonance, but are captured into the 7:3 resonance. Once
the planets are locked in resonance, they begin to migrate inwards
together so as to maintain the 7:3 commensurability. However, 
once resonance locking occurs, the amplitude of the secular eccentricity
oscillations between the two planets begins to increase, and the 
eccentricity of the smaller outer planet reaches a value of $\sim 0.2$
within 30,000 years.
In a simple N-body calculation, which does not
include the back reaction of the exterior gas disk, the outer planet
eccentricity 
grows to a point where the system is destabilized.  Lee and Peale (2002)
have addressed this problem by introducing parameterized eccentricity damping,
but fully self-consistent hydrodynamical calculations of the disk response
to the resonant migration of the planets need to be done.

The best-fit to the radial velocity data for 47 UMa indicates that
the system may indeed be in the 7:3 resonance, in which case eccentricity
damping is likely to have occured unless the nebula dissipated immediately
after the system was captured into resonance. A longer time baseline
of observations will nail down the period of the outer planet more 
accurately, and further simulations will be done to better elucidate
the process of eccentricity damping for planets caught in resonance.

In order to gain a better understanding of whether capture into the 7:3
resonance is likely, we have performed a large number of additional three-body
integrations in which a constant azimuthal acceleration is applied to the
outer planet. For systems in which the planets have initial osculating
elements corresponding to those in Table 1, but with $P_{c}>3000 {\rm d}$,
capture into the 3:1 resonance occured in every case we examined. 
We have also computed 10,000
simulations in which the outer planet starts with $P_{c}=2800 {\rm d}$
(i.e. interior to the 3:1 resonance), and
the initial osculating eccentricity of planet c is chosen randomly to
fall between 0.0 and 0.15. The results of these simulations are shown in
Figure 8.

This figure indicates that capture into the 7:3 resonance is possible, but
is much less likely than capture into the 5:2 resonance, which occurs
in about 50\% of the cases. (Note that many of the symbols in Figure 8
overlie one another at the 5:2 resonance.) 
In the simulations where the planets avoid
being caught in either the 5:2 or the 7:3, capture into the lower order 2:1
commensurability appears to be certain. It is clear that evolution
into the locked 5:2 or 7:3 commensurabilities depends
sensitively on the initial conditions for the integration. When the
eccentricity of the outer planet starts with a small value (which would
be expected from a standard gap-opening scenario) capture into the 7:3
is observed in 3\% of the simulations in which $e_{c_{i}}<0.02$.
For large values of the initial eccentricity for the outer
planet, capture into higher-order resonances (e.g. 9:4, 12:5, etc) is
occasionally observed. Note that capture into the 7:3 resonance for
small values of $e_c$ requires $e_b \sim 0.05$. When $e_b \sim 0.01$,
capture into the 2:1 resonance is certain for $e_c<0.04$.
All of the simulations resulted in some form of 
resonance capture, which suggests that pairs of resonant planets may
be quite common in the galaxy.

Although no planetary systems have yet been observed in the 5:2 resonance,
this configuration may be particularly ubiquitous. Indeed, Lee \& Peale (2002)
have suggested that Jupiter and Saturn may once have been participating 
in a 5:2 resonance, which was subsequently disrupted, perhaps as a result
of interactions with a remnant planetesimal or gas disk in the early Solar
System.

There is still some uncertainty in the period ratio of the 47 UMa planets,
but the system is clearly interior to the 3:1 resonance. Because capture into
3:1 resonance is virtually certain within our migration scenario, it appears
that the planets must have started interior to the 3:1 commensurability.

\section{Accretion of Earth-Like Planets}

In the Solar System, terrestrial planets formed and survived in a
large stable zone between 0 and 2 \au\ from the Sun which happens
to include the Sun's habitable zone (Kasting \etal\ 1993). The size
and width of this stable zone is largely determined by the orbits
and masses of the giant planets. Most orbits with semi-major
axes between 4 and 36 \au\ are unstable due to close encounters
with Jupiter, Saturn, Uranus or Neptune (\eg\ Gladman and Duncan
1990). Many orbits in the asteroid belts (roughly 2--4 \au\ and
$>36$ \au) are stable, but these regions contain a number of
unstable orbital resonances associated with the giant planets,
and it appears that these resonances prevented the formation and/or
survival of terrestrial planets in these parts of the Solar System
(\eg\ Franklin and Lecar 2000, Nagasawa \etal\ 2000, Chambers and
Wetherill 2001).

The 47 UMa system contains 2 giant planets with similar mass and
period ratios to Jupiter and Saturn, and orbits with low
eccentricities---a characteristic also shared by the giant planets
of the Solar System. Although the absolute masses and relative
inclination of the planets in the 47 UMa system are not known at present,
it is plausible that these giants could have sculpted a system of asteroid
belts and terrestrial planets similar to the Solar System, but shifted
to smaller semi-major axes reflecting the shorter orbital periods
of the giants.

Jones \etal\ (2001) have examined the stability of Earth-mass planets
in the presence of the inner giant (\ie\ before the existence of the
outer giant had been established). These authors found that for $\sin
i=1$, orbits inside $\sim 1.3$ \au\ are stable apart from a narrow
unstable zone at the 3:1 resonance at 1.0 \au.  For $\sin i=0.5$,
orbits are stable inside $\sim 1.15$ \au, and the unstable zone at the
3:1 is wider. In both cases, stable orbits can exist in a part of the
star's habitable zone. Chambers (2000) made a single simulation of
late-stage terrestrial-planet accretion in the presence of the inner giant
planet. In this simulation, several small terrestrial planets formed,
including a Mars-size body moving on a stable orbit at $a=0.92$, at
the inner edge of the habitable zone.

The presence of the outer giant complicates the dynamics of
terrestrial planets. In particular, a strong secular resonance,
analogous to the $\nu_6$ resonance in the Solar System, is present
at $a\sim 0.85$ \au, and any terrestrial planet forming in this region
will have an unstable orbit. (In the Solar System, the $\nu_6$ resonance
marks the inner boundary of the main asteroid belt.) We have examined
the stability of test particles in the presence of the two giant
planets using N-body integrations. The top panel of Fig.~9 shows
the fates of 280 particles with low-$e$, low-$i$ orbits and semi-major
axes $0.4<a<2.0$ \au, integrated for 20 million years using a
hybrid-symplectic algorithm (Chambers 1999). In the figure, blue
vertical bars indicate the initial semi-major axes of particles that
were ejected from the system by close encounters with one of the giant
planets. Yellow bars indicate particles that came within 0.1 \au\ of
the star, and we assume that these particles would ultimately be
removed by colliding with the star. Red bars indicate particles that
survived for the length of the integration.

Most of the test particles with $a>1.3$ \au\ are removed, mainly by
ejection onto hyperbolic orbits. This is similar to the result of
Jones \etal\ for the case with just the inner giant planet.
Many of the particles with $a<1.3$ \au\ survived, although there are a
number of narrow unstable regions associated with resonances with the
giant planets. The secular resonance at $a\sim 0.85$ \au\ removed
several particles by inducing large orbital eccentricities so that the
particles fell into the star.  Gaps at the 3:1 and 5:2 mean-motion resonances
are also apparent. Two particles survived in the 3:2 resonance at
$a\sim 1.6$ \au, although it is unclear whether they would survive
on longer timescales. In many ways the distribution of stable and
unstable regions mimics that in the main asteroid belt of the Solar System,
with Kirkwood gaps at the $\nu_6$, 3:1, 5:2 and 2:1 resonances,
and the Hilda asteroids in a stable region at the 3:2 resonance.

Most of the test particles with $a<1.3$ \au\ survived in the
simulation. However, this doesn't provide a robust indication of
whether terrestrial planets can form or survive here. Terrestrial
planets have mass and so they modify the strength and location of
resonances associated with the giant planets, as Earth and Venus
modify the $\nu_5$ resonance in the Solar System (Namouni \& Murray
1999).  During the late stages of planetary accretion, many planetary
embryos with appreciable mass will be present. These will modify
one another's orbits during close encounters as well as causing
secular orbital evolution due to more distant perturbations. 
In particular, embryos can gravitationally
scatter one another into unstable resonance regions, and this greatly
enhances the ability of resonances to remove material from a
protoplanetary disk (Wetherill 1992). Chambers and
Wetherill (2001) have shown that if planetary embryos formed in the
asteroid belt of the Solar System, there is a two thirds
probability that a combination of gravitational scattering and 
giant-planet resonances would have removed all embryos from the belt. 
Something similar may have occurred in the region beyond 0.85 \au\ 
in the 47 UMa system.

To examine this possibility, we have made 4 N-body simulations of
terrestrial-planet accretion in the 47 UMa system. We assume that
planetary embryos were able to form rapidly by runaway accretion
(Wetherill and Stewart 1993) interior to the orbits of the giant
planets. In addition, since the formation/migration timescales are not
known for the giant planets, we make the simplifying assumption that
the giants had attained their current orbits and masses at the start
of the final stage of accretion of the terrestrial planets.  To
calculate the orbital and accretional evolution of the embryos we used
a hybrid-symplectic N-body integrator (Chambers 1999).

Simulation 1 began with 280 lunar-mass bodies ($m\sim 3\times 10^{-8}
M_\odot$) with $0.3<a<2.0$ \au\ and roughly circular, coplanar orbits.
The total initial mass of the embryos is about 2.5 Earth masses, which
in the Solar System is generally sufficient to form a system of 3 or 4
planets with total mass similar to the terrestrial planets (Chambers
2001). In this simulation, the giant-planet masses are assumed to
have their minimum possible values (\ie\ we take $\sin i=1$.)
Figure~10 shows 6 snapshots of the evolution in the simulation.
Each panel shows the instantaneous values of $a$ and $e$ for surviving
embryos, with symbol radius proportional to the radius of the body.
Within $10^5$ years, the inner giant planet has gravitationally
ejected most of the embryos with $a>1.4$ \au, and a gap has developed
at the secular resonance at $a=0.85$ \au. After 0.5 million years,
the secular resonance is almost empty and embryos are also being
dynamically excited and removed at the 3:1 resonance near 1 \au.

By 2 million years, a large fraction of the initial population of
embryos has been lost, while little accretion has taken
place. Clearly, the timescale to remove material from the unstable
parts of the disk is short compared to the accretion timescale. It is
also apparent that depletion of material is not confined to the
resonance locations, but extends across most of the disk beyond about
0.7 \au. As a result, appreciable accretion is confined to the
innermost part of the disk which represents a large contiguous region
where resonances are absent. At the end of the simulation, this region
contains 3 small terrestrial planets, the largest of which has $m\sim
0.2$ Earth masses. All of these planets lie inside the inner edge of
the habitable zone of 47 UMa. A single unaccreted (lunar mass) body
survives in the habitable zone (HZ) at $a=1.2$ \au. The mass of this 
object simply reflects the initial mass of the embryos used in the 
calculation, and as such the size of this ``planet'' is not very meaningful.

The middle panel of Figure 9 shows the fates of the 280 embryos
at the end of Simulation 1, using the same colour coding as the
upper panel. In cases where 2 or more embryos merged, each of the
embryos is assigned the same fate as the composite body. Note that
the number of surviving objects is much smaller than in the
test-particle integration. In particular, embryos are lost in a wide
region with $a>0.5$ \au, and not merely at resonance locations. 

If planetary embryos did form in the HZ of 47 UMa, it is unlikely
that they would have had a uniform mass. In fact, runaway accretion
of planetesimals, which is thought to lead to the formation of embryos,
generally produces a few large objects and many smaller ones (Kokubo
and Ida 1998). With this in mind, we made a second accretion simulation which
began with 14 Mars-sized bodies ($m\sim 3\times 10^{-7} M_\odot$)
and 140 lunar-mass objects. The total mass and orbital distributions
were the same as in Simulation 1. Planetary accretion simulations
are notorously stochastic, so we were somewhat surprised that the
orbital and accretional evolution in Simulation 2 was very
similar to the first. The simulation ended with 3 small planets
with $a<0.7$ \au, with masses $m<0.2$ Earth masses, in addition to
a single Mars-sized embryo with $a\sim 1.2$ \au\ which had undergone
no accretion.

In Simulations 1 and 2, the total mass of embryos in the region
0--2 \au\ is similar to that which would have existed in a solar
minimum-mass nebula (Weidenschilling 1977). However, the giant planets
of 47 UMa are more massive than Jupiter and Saturn. In addition,
the giants probably migrated inwards after they formed, which suggests
that 47 UMa had a more massive protoplanetary disk than the Sun.
For this reason, we reran Simulations 1 and 2 enhancing the initial
masses of the embryos by a factor of 5 (the giant-planet masses were
unchanged). In each of these simulations, the dynamical evolution
progressed through the same stages seen in Simulations 1 and 2.
However, the embryos were typically removed somewhat more quickly
since their mutual gravitational perturbations were larger, and they
were scattered into unstable resonances more easily. Accretion was
again confined to the region $a<0.7$ \au, and in these simulations
the larger embryos produced fewer and more massive final planets.
Simulation 3 (uniform-mass embryos) ended with 2 approximately
Earth-mass planets with $a\sim 0.4$ and 0.7 \au. The individual 
fates of the initial bodies in Simulation 3 are shown in the bottom
panel of Figure 9. Again, objects are removed from a much wider
range of orbits than the test-particle case.

Simulation 4
ended with a large, 2-Earth-mass planet with $a\sim 0.5$ \au, and a
second planet with $m\sim 0.5$ Earth masses, and $a=1.1$ \au. In
common with Simulations 1 and 2, Simulation 4 ended with a single
body in the HZ which had undergone no accretion, so that the final mass
of this planet reflects the initial masses of the embryos. However,
we note that if large embryos did form during the runaway accretion
stage, a large unaccreted embryo could exist in the HZ of 47 UMa,
and this body may be large enough to support plate tectonics (as
Earth does, but Mars does not) and be considered a habitable planet.

\section{Discussion}

The discovery of a second planet orbiting 47 UMa raises a series
of very interesting dynamical questions, and can serve to provide 
important constraints on the formation history of the system.

The 47 UMa system is different from the other known multiple-planet
extrasolar systems in two important respects. First, both planets lie at
considerable distance from the star. The period of the inner planet,
47 UMa B is five times longer than the inner planet period of any
of the other known multiple systems (currently GJ876, Upsilon Andromedae,
HD 168443, HD83443, and HD 82943), and 47 UMa C has the longest period
of any known extrasolar planet. Second, the orbits of both planets 
are nearly circular, again in stark contrast to the other known
multiple-planet systems. Indeed, 47 UMa is currently the best 
extrasolar analog of our own system.

47 UMa B and C are engaged in a secular exchange of eccentricities with 
a period of roughly 6,000 years. A first-order theory does a good
job of describing this motion, and verifies that the two planets are
stabilized by a secular resonance which maintains the periapses of the
planets in libration around an aligned configuration. This secular
resonance is also observed in the outer two planets of Upsilon
Andromedae, and is likely a very common state of affairs in the Galaxy.
Jupiter and Saturn currently lie in a different region of secular
phase space and do not participate in the secular resonance, although
these planets would lie in the resonance for smaller values of
$e_{\rm SAT}$ or $\Delta\varpi$.

The current time baseline of radial velocity data covers only two full
periods of 47 UMa C, and non-Keplerian interactions between the two planets 
cannot be resolved with 3--5 m/s precision. The individual and mutual
inclinations are therefore unconstrained by the radial velocity measurements.
Dynamical calculations show that the system can tolerate mass factors of 
$\sin(i)=0.2$ for co-planar configurations. However, if the system
is non-coplanar, stability requires that the mutual inclination be 40 degrees
or less, and that this value is quite insensitive to the masses of the two
planets.

The small eccentricities of the two planets in the 47 UMa system 
put interesting constraints on the theory of planet-nebula migration. Our
results show that when 2-planet systems undergo interaction with an exterior
circumstellar disk, the outer planet will migrate inwards until it becomes
trapped in a low-order mean motion resonance with the interior planet.
Generally, this occurs in the 2:1 resonance, as is the case with GJ 876,
but it appears that other resonances, including the 3:1, 7:3, and 5:2 can 
also elicit capture. Once two planets are captured in resonance, they
migrate inwards together, and their eccentricities increase. It is not
clear how eccentricity damping can occur to the degree that both planets
remain on nearly circular orbits.

It seems quite possible that the class of eccentric giant planets discovered
to date may be the end result of resonant migration with eccentricity pumping.
Once planetary orbits begin to cross, the smaller planet is generally
ejected from the system, leaving behind the larger survivor on an eccentric
orbit. This may explain the fact that eccentric giant planets have a
larger average mass than the short-period 51 Peg type planets, which 
presumably were able to migrate all the way in without being affected by
a second body. 

A possible scenario for 47 UMa might run as follows. The inner planet
formed in the outer part of the 47 UMa protoplanetary nebula, 
possibly beyond the ice-line, and
then migrated inwards towards its present location at 2 \au. During
this migration, a second planet may have formed as a result of density
enhancements or vortices beyond the inner edge of the exterior disk. The
growth of this second planet would have led to further gap opening, and
an isolation of the inner planet and a halt to its migration. The
remnant disk was then unable to force much further migration of 47 UMa c,
accounting for its present circular orbit. Had it migrated further,
47 UMa c would have been locked in the 2:1 resonance, would have had its
eccentricity pumped, and would have escaped, leaving 47 UMA b as just
another eccentric giant.

Earth mass planets can survive in the habitable zone of 47 UMa, but
accretion simulations show that it is unlikely that they could have
formed there unless very massive planetary embryos accreted in the
habitable zone prior to the formation of the giant planets. 

\acknowledgements
We thank Peter Bodenheimer, Man Hoi Lee, and Stan Peale for useful
advice. An anonymous referee provided helpful criticisms.
This material is based upon work supported by NASA under contract No.
RTOP 344-37-22-12 issued through the Origins of Solar Systems Program.
Many of the numerical calculations were performed using the Condor cluster
run by the NAS program at Ames Research Center. JEC acknowledges support
from an NRC Postdoctoral Program and from the NASA Origins of Solar 
Systems Program.

\clearpage

\begin{figure}
\figcaption{
Variations in eccentricity (left-hand side panels) and
relative apsidal alignment (right-hand side panels)
for (1) the best-fit system from Fischer et al (2002) (top panels),
(2) the best-fit system from Fischer et al (2002) with $e_2$=0.15
initially (middle panels),
and (3) for Jupiter and Saturn (bottom panels)
The black curves show the analytic Laplace-Lagrange secular
solution of the respective two-planet problems, and the red dots
show numerical integrations. Note that the $x$-axes are plotted
in units of inner planet orbits to facilitate comparison between
the two systems.
}
\label{fig1}
\end{figure}

\begin{figure}
\figcaption{
Libration regions for the planets in 47 UMa, and for Jupiter and Saturn, as
a function of the initial values of $\Delta\varpi$ and $e$ of the
outer planet.
}
\label{fig2}
\end{figure}

\begin{figure}
\figcaption{
Survival and destruction of mutually inclined two planet fits to the
47 UMa radial velocity data.
The results of million-year long dynamical integrations are organized
according to
the mutual inclination and the combined total planetary mass of each
two planet fit. Systems in which the outer planet is observed to reach
${e_2}_{max}>0.4$ are plotted as filled circles. Systems in which the outer
planet reaches $0.2< {e_2}_{max} <0.4$ are unlikely to remain stable for the 
7 Gyr age of 47 UMa, and are plotted as stars, whereas stable systems,
with ${e_2}_{max}<0.2$, are plotted as open circles.
}
\label{fig3}
\end{figure}

\begin{figure}
\figcaption{
Surface density distribution of the
disk after 250 orbits of the inner planet, after a steady 
state has been achieved.
}
\label{fig4}
\end{figure}

\begin{figure}
\figcaption{
Results of a hydrodynamical simulation of
two planets (with the nominal masses of 47 UMa B and C)
embedded in a protoplanetary disk.
The two planets start out beyond the 7:3 resonance, with a 
semi-major axis ratio $a_{2}/a_{1}$=1.8. The $y$-axes of the
second and third panels from the top are in units of 2 \au.
There is initially 1 \mjupe in the external gaseous
disk. The top panel of the figure shows that the disk exerts
a steady negative torque on the outer planet, but has little
coupling to the inner planet. The torque on the outer planet
causes the orbit of the outer planet to migrate steadily inward
(2nd from top), whereas the inner planet stays nearly fixed.
In this uncoupled migration phase, the eccentricities do not
increase (bottom panel).
}
\label{fig5}
\end{figure}

\begin{figure}
\figcaption{
Surface of section showing the results of a three-body integration
in which the torque measured in the hydrodynamical simulation shown
in Figures 4--5 is applied to the outer planet. The orbit of the
outer planet decays, skirts the 5:2 resonance, and lands in the
7:3 resonance. This simulation shows that capture in the 7:3 
resonance is possible.
}
\label{fig6}
\end{figure}

\begin{figure}
\figcaption{
Evolution of the three-body simulation with a steady azimuthal torque
applied to the outer planet. Capture into the 7:3 resonance occurs
approximately 6000 years into the simulation.
}
\label{fig7}
\end{figure}

\begin{figure}
\figcaption{
Results of 10,000 three-body simulations with steady azimuthal torque
applied to the outer planet. Systems which do not experience capture
into higher-order resonances are captured into the 2:1 resonance.
Capture into high-order resonances (listed on the right-hand margin)
are increasingly likely for large initial eccentricity of the outer
planet. 
}
\label{fig8}
\end{figure}

\begin{figure}
\figcaption{
{\it Upper:} the fate of test particles starting on nearly circular, co-planar
orbits integrated for 20 million years in the 47 UMa system. Red = survives,
blue = ejected, yellow = collision with central star. {\it Middle, lower:} the
fate of planetary embryos in two accretion simulations (embryo masses 1, 5
lunar masses respectively). Note that additional regions are rendered
unstable compared to the test particle case due to gravitational interactions
between embryos.
}
\label{fig9}
\end{figure}

\begin{figure}
\figcaption{
Six snapshots in time of a planetary accretion simulation in the 47 UMa system.
The simulation began with 280 lunar mass planetary embryos with semi-major
axes $0.4<a<2$ \au\  encompassing the star's habitable zone. Embryos are 
removed rapidly from much of this region as mutual gravitational perturbations
scatter them into unstable resonances associated with the giant planets.
Accretion is confined to the region $a<0.7$ \au\, inside the inner edge of
the habitable zone.
}
\label{fig10}
\end{figure}

\clearpage

\begin{table}
\tablenum{1}
\caption{Orbital Parameters}
\begin{tabular}{lrrrr}
\tableline
\tableline
Param          		   & 47 UMa (b)  &           & 47 UMa (c)             \\
\tableline								                                  
P (d)         		   & 1089.0 (2.9)         &  & 2594 (90)              \\
${\rm T}_{\rm p}$ (JD)     & 2453622.9 (33.6)     &  & 2451363.5 (495.3)      \\
e              		   & 0.061 (0.014)        &  & 0.005 (0.115)         \\
$\omega$ (deg) 	           & 171.8 (15.2)         &  & 127.0 (55.8)           \\
K$_1$ (\ms)    		   & 49.3 (1.2)           &  & 11.1 (1.1)             \\
a (\au)                     & 2.09                 &  & 3.73                   \\
a$_1 \sin i$ (\au)	   & $4.94 \times 10^{-3}$&  & $2.64 \times 10^{-3}$  \\
f$_1$(m) (M$_\odot$)       & $1.35 \times 10^{-8}$&  & $3.67 \times 10^{-10}$ \\
M$_2 \sin i$ (M$_{Jup}$)   & 2.54                 &  & 0.76                   \\       
$^1{\rm Nobs}$         &    & 90                     &                      \\
RMS (\ms)              &    & 7.4                    &                      \\
$\chi^2$               &    & 1.06                   &                     \\
\tableline
\end{tabular}
\end{table}
\clearpage

\centerline{\bf Table 2}
\medskip
\centerline{\bf Highest Eccentricity for ``c'' in 100 Myr}
\medskip
\begin{center}
\begin{tabular}{rrrrrrrrr}
\hline
$\,$ & $\,$  & $\,$ & $e_{c} (initial)$ & $\,$ & $\,$ & $\,$ & $\,$ & $\,$ \\
$\sin(i)^{-1}$  & 0.005 & 0.05 & 0.10 & 0.15 & 0.175 & 0.20 & 0.225 & 0.25 \\
\hline
1.000 & 0.099  & 0.097 & 0.127 & 0.172 & 0.198 & unst & unst & unst    \\
1.125 & 0.102  & 0.100 & 0.129 & 0.176 & unst & unst  & unst  & unst    \\
1.250 & 0.110  & 0.111 & 0.132 & 0.179 & unst & unst  & unst  & unst    \\
\cline{9-9}
1.500 & 0.124  & 0.123 & 0.154 & unst & unst & unst & unst & \multicolumn{1}{|r}{unst}  \\
\cline{8-8}
1.750 & 0.109  & 0.126 & unst & unst & unst & unst & \multicolumn{1}{|r}{unst} & unst  \\
\cline{7-7}
2.000 & 0.112  & 0.113 & 0.137 & unst & unst & \multicolumn{1}{|r}{unst} & unst & unst \\
\cline{5-6}
2.500 & 0.118  & 0.121 & 0.142 & \multicolumn{1}{|r}{unst} & unst & unst & unst & unst \\
\cline{4-4}
3.000 & 0.123  & 0.128  & \multicolumn{1}{|r}{0.149} & unst  & unst & unst & unst & unst \\
\cline{2-3}
3.500 & 0.126  & 0.133  & 0.156 & unst  & unst & unst & unst & unst \\
4.000 & 0.140  & 0.145  & 0.164 & unst  & unst & unst & unst & unst \\
5.000 & 0.128  & 0.138  & unst & unst  & unst & unst & unst & unst \\
6.000 & 0.133  & 0.127  & unst & unst  & unst & unst & unst & unst \\
\hline
\end{tabular}
\end{center}

\end{document}